\documentstyle[12pt,aaspp4_rj,psfig]{article}
\def\cge      {{$_ >\atop{^\sim}$}}
\def\cle      {{$_ <\atop{^\sim}$}}
\def\ztpf     {{$z\simeq 2.4$}}
\def\hst      {{\sl HST}}
\def\wfpc     {{\sl WFPC2}}
\def\etal     {{\it et\thinspace al.} }

\def\eg       {{\it e.g.}}
\begin{document}

\title{Compact Lyman-$\alpha$ Emitting Candidates at \ztpf\ in Deep Medium-band
\hst\footnote{Based on observations with the NASA/ESA {\it Hubble Space Telescope}
obtained at the Space Telescope Science Institute, which is operated by AURA, Inc., under
NASA Contract NAS 5-26555.} \wfpc\ Images}

\author{Sebastian M. Pascarelle}
\affil{Dept. of Physics \& Astronomy, SUNY Stony Brook, Stony Brook,
NY 11794}
\author{Rogier A. Windhorst}
\affil{Dept. of Physics \& Astronomy, Arizona State University, Tempe,
AZ 85287}
\authoremail{smp@deltat.la.asu.edu, raw@cosmos.la.asu.edu}
\and
\author{William C. Keel}
\affil{Dept. of Physics \& Astronomy, University of Alabama, Tuscaloosa,
AL 35487}
\authoremail{keel@bildad.astr.ua.edu}

\begin{abstract}
Medium-band imaging with HST/WFPC2 in the F410M filter has previously revealed
a population of
compact Lyman-$\alpha$ emission objects around the radio galaxy 53W002
at $z\simeq 2.4$. We report detections of similar objects at $z\simeq 2.4$
in random, high-latitude HST parallel observations of three additional
fields, lending support to the idea that they constitute a
widespread population at these redshifts. The three new fields contain 18
Lyman-$\alpha$ candidates, in contrast to the 17 detected in the
deeper exposure of the single WFPC2 field around 53W002. We find
substantial differences in the number of candidates from field to field,
suggesting that significant large-scale structure is already present in the
galaxy distribution at this cosmic epoch. The likely existence of
\ztpf\ sub-galactic clumps in several random fields shows that these objects
may have been common in the early universe and strengthens the argument that
such objects may be responsible for the formation of a fraction of the luminous
present-day galaxies through hierarchical merging.
\end{abstract}


\section{Introduction}

The formation of structure in the universe, from the largest scales ({\it
e.g.}, ``great walls'') to the smallest scales ({\it e.g.}, galaxies and
globular clusters), remains a very controversial topic in modern observational
cosmology. Standard inflationary cosmologies predict that the primordial
density field is a Gaussian random field which can lead to either ``top-down''
or ``bottom-up'' structure formation, depending on whether very small or very
large structures formed first. This in turn depends upon whether the
gravitationally dominant dark matter is cold (CDM), hot (HDM), or
some mixture thereof (CHDM, MDM).

Traditionally, it has been assumed that a dominant dissipationless component of
dark matter accumulates through gravitational instability into clumps, the
virialized parts of which become dark matter halos. Galaxies then form by the
cooling and condensation of gas within these halos (White \& Rees 1978). The
inability to detect the intense Lyman-$\alpha$ emission expected from large
numbers of such young forming galaxies with strong continuous star formation
at high redshift in narrow-band imaging surveys (Thompson, Mannucci, \&
Beckwith 1996; Thompson, Djorgovski, \& Trauger 1995) is instead more
consistent
with a hierarchical formation scenario, in which galaxies were assembled
from pieces over a long interval. It is also possible that high-redshift
galaxies, or more specifically, their Lyman-$\alpha$ light is obscured by
dust or destroyed by resonant scattering, even in the absence of dust. Still,
hierarchical clustering (a ``bottom-up'' scenario) currently appears to be the
most successful at reproducing many of the properties of the real universe,
as noted by Diaferio \& Geller (1997), while remaining within the constraints
produced by the COBE results. In the typical (CDM) hierarchical
clustering model, structure grows from the gravitational collapse of small
initial density perturbations, from which systems of progressively larger mass
merge and collapse to form newly virialized systems. Most modern hierarchical
models are variations of the original Press-Schechter (1974) theory of structure
formation, in which isothermal ``seeds'' existing at
recombination grew into the galaxies we see today.

Pascarelle \etal (1996b, hereafter P96b) presented evidence for a 
gravitationally bound collection of sub-galactic star-forming objects
at \ztpf, which fits well with expectations from hierarchical galaxy
formation. These objects appear to have effective radii (or half-light
radii) of $r_e$\cle 0\farcs 1, or \cle 1 kpc (assuming $H_{\rm o}$=100 km
s$^{-1}$ Mpc$^{-1}$, $q_{\rm o}$=0.5 throughout, unless otherwise noted).
Since this is much smaller than the median \wfpc\ field galaxy scale length of
$\simeq$0\farcs 3 (Odewahn \etal 1996), and at any redshift in the range
$z\simeq 0.3-3$, 0\farcs 1--0\farcs 2 corresponds to only 0.5--2 kpc, the
\ztpf\ candidates appear to be smaller than the size of a present-day
galaxy ({\it i.e.}, sub-galactic). P96b suggested that these proposed ``galaxy
building blocks'' are actually part of a widespread population existing
throughout the redshift range $z\simeq 2-5$. We
undertook a Cycle 6 \hst\ project to image four random fields using
the same filters as P96b (since WFPC2 conveniently has a medium-band filter
to sample Lyman-$\alpha$ at \ztpf) in an effort to test the universality of
these objects at \ztpf.

It is clear that galaxies at all redshifts are more inclined to be in
groups or clusters than isolated, giving the term ``field galaxy'' less
meaning than it once had. As such, it is perhaps not surprising to find
that in many ``pencil-beam'' galaxy redshift surveys, often several galaxies
are found in a single narrow redshift bin ({\it e.g.}, the Hubble Deep Field
(HDF) counts reported in Cohen \etal 1996 and the large structure of galaxies
found at $z\simeq 3$ by Steidel \etal 1998). On a slightly larger angular
scale, ``spikes'' have been found in the redshift distribution
of several more extensive galaxy surveys out to $z$\cle 1 (Broadhurst \etal
1990; Landy \etal 1996; Ettori, Guzzo \& Tarenghi 1997),
indicative of some sort of larger structure at earlier epochs. What is
questionable, according to CDM models, is whether or not any such structure
could have existed at high redshifts. Rauch, Haehnelt, \& Steinmetz (1997,
hereafter RHS97) are able to reproduce
the observational characteristics of QSO absorption systems at $z\sim 3$ over a
wide range of column densities with a model consisting of sub-galactic
clumps embedded in sheet-like structures and often lying along
filaments. The similarities between the grouping of \ztpf\ sub-galactic
clumps of P96b (see their Fig. 1) and the RHS97 simulations are quite striking
(see Fig. 1{\it a} of RHS97, which shows not only a field about the same size
as our 53W002 \wfpc\ field, but also with about the same number of clumps at a
similarly high redshift). Their simulations indicate that $\sim$20 compact
clumps will merge to produce about three $L^*$ galaxies by $z$=0. This agrees
quite well with the calculations carried out in P96b, which suggest that 17
candidate subgalactic clumps could form several $L^*$ galaxies by $z$=0. Is it
possible that groups, clusters,
or even larger-scale structure existed to some extent at earlier epochs,
although at much lower amplitude than that seen at lower
redshifts? If the answer is yes, then it should not be surprising to find
the compact, star-forming protogalactic objects arranged in groups, clusters,
or even filaments or ``sheets.'' The Cycle 6 data presented here represent
an initial attempt to begin to answer this question.

In \S 2, we describe the observations and data processing, in \S 3 we
present the results, and in \S 4 we discuss some of the implications of these
results in the context of galaxy and structure formation at high redshift.
Our conclusions are summarized in \S 5.

\section{Observations}

\hst\ has the ability to image a field with \wfpc\ in parallel with other
instruments. This (usually) random field can be 4\arcmin--14\arcmin\ away from
the primary
target, depending on the primary instrument being used. We received three long
parallel observations with \wfpc\ using the F410M, F450W, and F814W filters
in Cycle 6. The medium-band F410M filter is centered at 4090\AA\ with a
full-width-at-half-maximum of 150\AA, so that Lyman-$\alpha$ emission from
objects in the range $z\simeq 2.30 - 2.42$ can be detected. We received
6--8 parallel orbits per field, divided where possible into more than one
exposure to allow proper cosmic ray removal (see Windhorst \etal 1994b). The
details of the observations,
including the dates, coordinates, position angle of the V3 axis, and total
exposure times in each filter are given in Table 1.

\begin{table}[h]
\caption{Cycle 6 Parallel Observations}
\begin{tabular}{cccccccc}
\tableline \tableline
date & $\alpha$\tablenotemark{a} & $\delta$ & PA of V3 & t$_{exp}$(F410M) & t$_{
exp}$(F450W) &
t$_{exp}$(F814W) \\
\tableline
30 Aug 1996 & 21$^{\rm h}$ 7$^{\rm m}$ 31\fs 8 & --5\arcdeg 22\arcmin 33\farcs
1 & 275\fdg 0 & 9600s & 2900s & 700s \\
11 Jan 1997 & 16$^{\rm h}$ 36$^{\rm m}$ 38\fs 3 & +82\arcdeg 34\arcmin 13\farcs
6 & 127\fdg 3 & 11900s & 3200s & 900s \\
20 May 1997 & 10$^{\rm h}$ 24$^{\rm m}$ 38\fs 1 & +47\arcdeg 4\arcmin 39\farcs
6 & 290\fdg 0 & 14900s & 5600s & 3000s \\
\tableline
\end{tabular}
\tablenotetext{a}{Coordinates are J2000 and are for the \wfpc\ field center}
\end{table}

The images were processed by the usual \hst\ pipeline, and aligned to the
nearest integer pixel in IRAF. For accurate photometry of compact objects
that are expected to have sizes of only a few \wfpc\ pixels or less, it is
imperative to do reliable cosmic ray rejection. Because of this, the averaging
and cosmic ray rejection were done with a custom-written code created to deal
specifically with a small number ($n\le 4$) of low-signal images.
This routine, written in the Interactive Data Language (IDL), first scales the
input images by their respective exposure times. Any remaining differences
left in the global sky values are from time-dependent variations in the sky
background. These variations are likely due to the proximity of any particular
image pointing to the earth's limb, which changes during each orbit, so that
the sky differences
are additive. The program calculates rough sky values from the global image
modes, and applies zeropoint offsets (by subtracting a constant) to bring all
the sky values to the lowest sky in the normalized set.
The image mode is calculated by fitting a polynomial to
the histogram of pixel values and finding the maximum of the curve
({\it i.e.}, the mode). Because the cosmic ray rejection code assumes all
pixels values within a few sigma of sky are ``good,'' it is important to
remove all bad negative pixels first. Therefore, all negative pixels more than
3.5$\sigma$ below the median of the eight surrounding pixels are replaced by
that median.

The main loop to remove cosmic rays works by ``stacking'' the aligned images
on top of
each other and then working on each individual ``stack'' of pixels (or array
of pixels in the $z$-dimension), after each pixel stack is sorted by value
in ascending order. The two pixels with the lowest
values are compared to see if they are $<$2.5$\times\sigma$ apart, where
$\sigma$ is the standard deviation using the lowest pixel value.
If so, then they are averaged, and the next highest value pixel is compared to
this average, and so on for the entire pixel stack. If a pixel has a value
$>$2.5$\times\sigma$ from this running average, then it is rejected as a
cosmic ray. The cutoff value of 2.5$\times\sigma$ comes from Windhorst,
Franklin, \& Neuschaefer (1994b), who determined this as the optimal level at
which to reject cosmic rays from multi-orbit {\sl WF/PC} images. The standard
deviation ($\sigma$) of the running average is
recalculated at each step for each pixel stack using the CCD equation:
\begin{displaymath}
\sigma=\sqrt{Sg+RN^2+Dt}
\end{displaymath}
in which S = the intensity (in ADU) of that pixel, g = the CCD gain
(in electrons/data number), RN = the CCD readnoise (in electrons), D = the CCD
dark current (in electrons/second), and t = the exposure time (in seconds).
The values for the gain, readnoise, and dark current were obtained from the
\wfpc\ handbook.

Photometry was then performed, as detailed in Windhorst \etal (1991), on the
cleaned average images using the CASSANDRA photometry package of D. Schneider,
which utilizes user-input apertures and sky regions. The sky region is
interpolated underneath the object aperture by fitting a sloped plane to the
pixels unaffected by faint neighbors (see also Odewahn \etal 1996). The
photometric zeropoints used are those
given in Holtzman \etal (1995) and include the various gain
ratios for each chip ($\simeq$19.55 for F410M, $\simeq$21.92 for F450W, and
$\simeq$21.59 for F814W). A total of 141 objects were detected simultaneously
in all three filters in the three fields. The photometric ``curve of growth"
was examined, and apertures were adjusted to give the highest signal-to-noise
ratio for each object, while maintaining the {\it same size} aperture in
the three filters.

\section{Results}

\subsection{New \ztpf\ Candidates}

The results of the photometry performed on the three Cycle 6 \hst\ fields are
presented in the color-color diagram of Figure 1 (the same as that used to
identify the \ztpf\ candidates of Pascarelle \etal (1996a) and P96b). For
consistency with the Cycle 5 53W002 field data, we interpolated $B_{\rm
F450W}-V_{\rm F606W}$ colors from the $B_{\rm F450W}-I_{\rm F814W}$ colors
using the effective central wavelengths (from the \wfpc\ handbook) of the
three filters involved.
This approximation is justified as long as the spectra of these faint blue
objects are close to a power law across the three filters (see Windhorst \etal
1991; P96b) and affects only the horizontal axis. As such, this approximation
has almost no effect on the nearly horizontal distribution of points and the
number of Lyman-$\alpha$ emitting candidates found.

\begin{figure}
\vspace*{-5cm}
\psfig{file=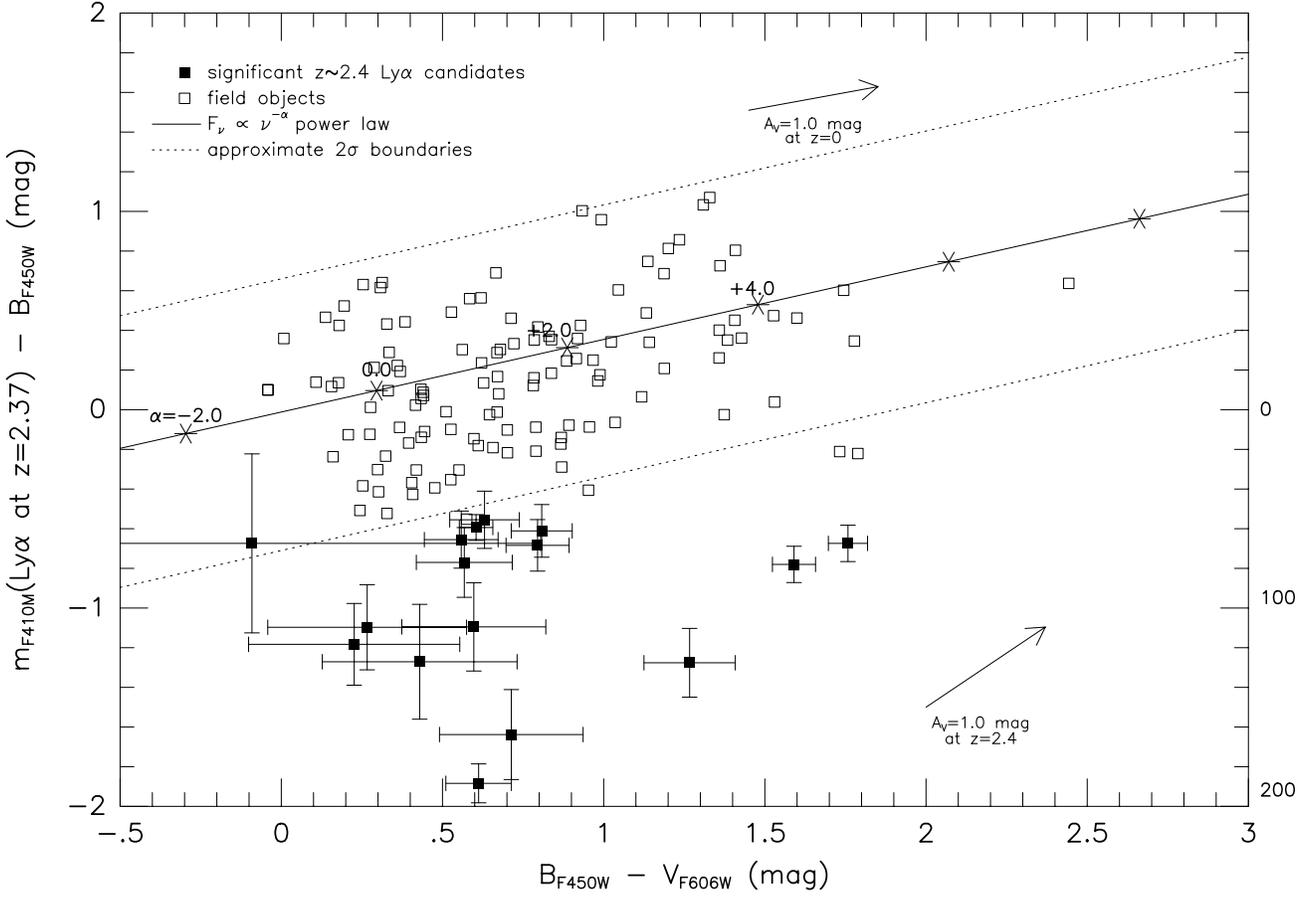,width=15cm}
\caption{($m_{\rm F410M}$--$B_{\rm F450W}$) versus ($B_{\rm F450W}$--$V_{\rm
F606W}$) color-color diagram for
145 objects detected in the three Cycle 6 fields. The open squares are field
objects and are grouped around the featureless power-law line. The
dotted lines represent approximate 2$\sigma$ errors in the photometry, and
contain most of the field objects. Below the lower boundary is a population
of objects, represented by filled squares with errorbars, which have excess
emission in the F410M filter. As described in the text, most of these
candidates are likely Lyman-$\alpha$ emitters at \ztpf. Unfilled squares
below the lower 2$\sigma$ boundary are candidates that were rejected as
stars by inspection. Approximate restframe equivalent widths are indicated
along the righthand axis, and show that most candidates have
W$_{\rm Ly\alpha}<100$\AA.}
\end{figure}

A featureless power law (F$_{\nu}\propto \nu^{-\alpha}$) is indicated by the
solid line, with asterisks indicating integer values of $\alpha$. Because of
the relatively small range of the restframe spectral energy distribution (SED)
that the HST filters will sample below the 4000\AA\ break for objects at high
redshift, a power law provides an adequate fit to their SEDs for various
values of $\alpha$. Since a large fraction of high-redshift objects are blue
and have relatively uncomplicated SEDs of blue stars shortward of the 4000\AA\
break, an $\alpha$ of $\sim$2 would be appropriate
to describe the Rayleigh-Jeans tail of their SEDs. Indeed, most of the objects
in Figure 1 group themselves around a value of $\alpha=0-2$ on the
power-law line. Lower redshift (older) stellar populations will appear redder
in these passbands, both from their intrinsic SED shapes and from the
$K$-corrections, and thus have larger values of $\alpha$.
The two dotted lines represent approximate 2-$\sigma$ photometric error
boundaries and mark the limits
of the general ``field" object distribution (open squares). Below this field
distribution on the plot lies a group of objects (filled squares) which have
significant excess emission at 4100{\AA}, likely Lyman-$\alpha$ emission at
\ztpf.

There exists the possibility that a fraction of these candidates could be
$z\simeq 0.097$ objects whose [OII] emission at 3727\AA\ has entered the
F410M filter. However the differential volume element at
\ztpf\ is $\sim$100 times larger than at $z\simeq 0.097$, and any other
emission lines between Lyman-$\alpha$ and [OII] strong enough to be detected,
such as \ion{C}{4} or \ion{Mg}{2},
would only come from objects that have much lower surface densities ({\it
e.g.}, QSOs and/or powerful radio galaxies).
Since no other strong emission lines exist between Lyman-$\alpha$ and [OII]
for star-forming objects (Kinney \etal 1996), we therefore expect that at most
a few of the candidates may turn out to be \ion{O}{2} emission line galaxies
(or others) at lower redshift.
Approximate Lyman-$\alpha$ restframe equivalent widths were calculated from
the F410M fluxes
and are indicated along the righthand vertical axis of Figure 1. If the excess
4100\AA\ emission is interpreted as Lyman-$\alpha$ emission, it can be
seen that most candidates have Lyman-$\alpha$ equivalent widths more
consistent with typical ``case B" recombination ({\it i.e.}, $<$100{\AA})
rather than with the existence of an AGN. The restframe ultraviolet reddening
vector (Seaton 1979)
expected at \ztpf\ is indicated, and suggests that the few reddest candidates
may have a visual absorption of $A_{\rm V}$\cle 2--3 mag. These reddest
candidates should be viewed with some caution, however, because of the slight
dependence of the $B_{\rm F450W}$ magnitudes on $B_{\rm F450W}$--$V_{\rm
F606W}$ color. This effect increases towards redder color (Holtzman \etal
1995), in the sense that these very red objects appear somewhat lower on the
plot than they should, and therefore could be contaminating field objects.

\begin{figure}
\vspace*{-5cm}
\psfig{file=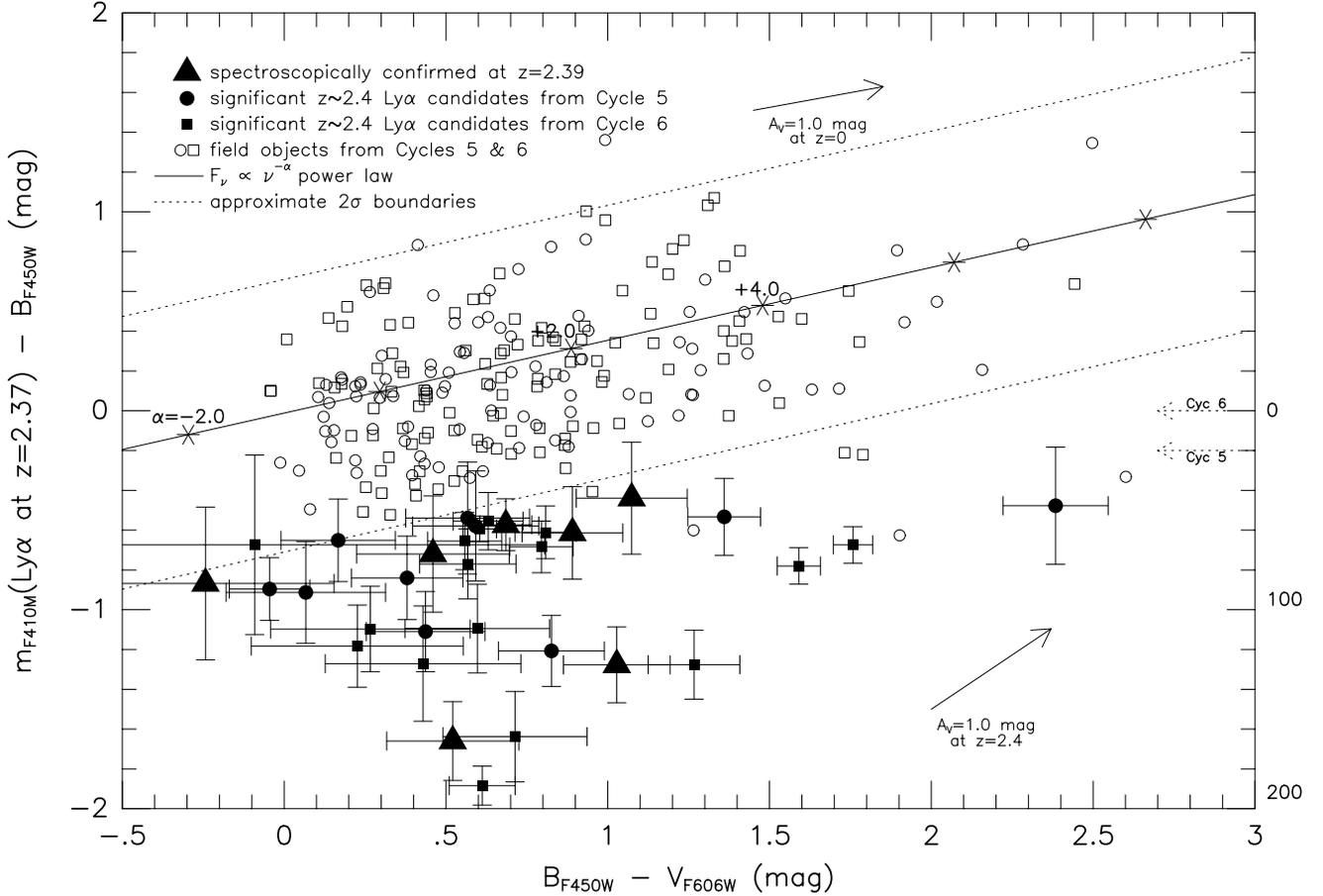,width=12cm}
\caption{The same ($m_{\rm F410M}$--$B_{\rm F450W}$) versus ($B_{\rm
F450W}$--$V_{\rm F606W}$) color-color diagram as in Figure 1, this time
containing all objects from Cycles 5 and 6. The squares (open and filled) are
from the three Cycle 6 parallel fields, and the circles (open and filled) are
from the Cycle 5 observations of the 53W002 field. Filled symbols are \ztpf\
candidates, and the filled triangles are objects from Cycle 5 that have been
spectroscopically
confirmed to be at \ztpf. Note the consistency of the photometry in that both
sets of field objects fall on top of each other. The median ($B_{\rm
F450W}$--$V_{\rm F606W}$) colors of the candidates are 0.57 mag and 0.60 mag
for the Cycle 5 and Cycle 6 data respectively. The two dotted arrows on the
right side of the figure represent the approximate completeness limits imposed
by the F410M observations for Cycles 5 and 6.}
\end{figure}

\begin{table}
\caption{Cycles 5 and 6 \ztpf\ Candidates}
\scriptsize
\begin{tabular}{ccccccccc}
\tableline \tableline
No. & $\alpha$(J2000) & $\delta$(J2000) & $B_{\rm F450W}$ & $V_{\rm F606W}$ & $I_{\rm F814W}$
& EW(Ly$\alpha$)\tablenotemark{a} & $r_{\rm e}$\tablenotemark{b} & $z$ \\
\tableline
1 & 17:14:10.53 & 50:15:09.80 &  26.0 & 24.5 & 24.2 & 35.0 & 0.16 & - \\
5 & 17:14:13.54 & 50:15:34.73 &  24.4 & 24.3 & 24.0 & 66.5 & 0.15 & - \\
6 & 17:14:14.74 & 50:15:28.91 &  23.4 & 22.6 & 21.9 & 38.0 & 0.72 & 2.390 \\
8 & 17:14:06.83 & 50:16:02.60 &  26.2 & 25.6 & 25.5 & 49.9 & 0.10 & 2.386 \\
11 & 17:14:12.63 & 50:15:36.75 &  25.6 & 24.5 & 24.0 & 28.4 & 0.30 & - \\
12 & 17:14:11.66 & 50:15:49.36 &  25.2 & 24.2 & 23.7 & 41.2 & 0.24 & 2.388 \\
13 & 17:14:10.98 & 50:15:54.26 &  26.3 & 23.9 & 22.0 & 31.0 & 0.12 & - \\
18 & 17:14:11.89 & 50:16:00.51 &  23.6 & 23.0 & 22.3 & 179.6 & - & 2.393 \\
19 & 17:14:11.27 & 50:16:08.72 &  24.1 & 23.0 & 22.2 & 113.0 & 0.36 & 2.397 \\
29 & 17:14:18.89 & 50:16:21.24 &  27.1 & 27.3 & 27.2 & 63.7 & 0.16 & - \\
34 & 17:14:13.44 & 50:16:57.08 &  25.1 & 24.8 & 24.3 & 44.1 & 0.20 & - \\
37 & 17:14:20.24 & 50:15:50.16 &  25.5 & 25.4 & 24.4 & 68.2 & 0.18 & - \\
40 & 17:14:20.76 & 50:15:41.83 &  25.5 & 25.0 & 24.0 & 90.6 & - & - \\
60 & 17:14:08.16 & 50:15:52.76 &  25.8 & 25.2 & 24.9 & 38.4 & 0.28 & - \\
61 & 17:14:08.53 & 50:15:52.22 &  26.3 & 25.7 & 25.4 & 35.5 & 0.17 & - \\
94 & 17:14:21.98 & 50:15:36.75 &  24.9 & 24.0 & 23.3 & 103.2 & - & - \\
113 & 17:14:16.84 & 50:16:00.16 &  25.7 & 25.2 & 24.4 & 60.9 & 0.08 & -\\
213 & 21:07:32.11 & -05:22:02.59 & 22.9 & - & 21.4 & 13.8 & 0.07 & - \\
222 & 21:07:30.25 & -05:21:29.86 & 24.1 & - & 19.8 & 19.3 & 0.12 & - \\
230 & 21:07:30.92 & -05:22:38.95 & 26.2 & - & 25.0 & 65.4 & 0.37 & - \\
301 & 16:36:29.45 & 82:34:45.65 & 27.2 & - & 27.1 & 477.0 & 0.09 & - \\
309 & 16:36:44.44 & 82:35:38.43 & 25.1 & - & 23.7 & 39.2 & 0.33 & - \\
314 & 16:36:42.33 & 82:34:47.00 & 26.4 & - & 24.6 & 163.9 & 0.31 & - \\
316 & 16:36:49.01 & 82:33:16.36 & 25.2 & - & 23.7 & 229.8 & 0.41 & - \\
321 & 16:37:08.00 & 82:33:56.15 & 24.3 & - & 20.4 & 55.9 & 0.28 & - \\
326 & 16:37:11.38 & 82:34:03.39 & 25.1 & - & 23.5 & 39.2 & 0.32 & - \\
337 & 16:36:13.27 & 82:33:31.59 & 25.2 & - & 23.2 & 39.2 & 0.52 & - \\
338 & 16:36:12.78 & 82:33:56.77 & 25.3 & - & 23.8 & 55.9 & 0.31 & - \\
339 & 16:36:18.22 & 82:33:53.24 & 25.7 & - & 22.6 & 113.8 & 0.14 & - \\
342 & 16:36:34.47 & 82:33:20.26 & 24.6 & - & 22.6 & 55.9 & 0.30 & - \\
343 & 16:36:04.83 & 82:34:06.17 & $>$27.8 & - & $>$27.5 & - & - \\
416 & 10:24:38.27 & 47:03:23.18 & 25.8 & - & 25.2 & 87.4 & 0.32 & - \\
422 & 10:24:39.87 & 47:03:24.74 & 26.6 & - & 26.7 & 47.1 & 0.23 & - \\
450 & 10:24:44.37 & 47:05:33.81 & 25.8 & - & 25.1 & 87.4 & 0.14 & - \\
459 & 10:24:39.74 & 47:05:56.63 & 25.9 & - & 24.4 & 87.4 & 0.25 & - \\
\tableline
\end{tabular}
\tablenotetext{a}{Estimated from the F410M and F450W fluxes}
\tablenotetext{b}{Mean of effective radius measured in each available broadband
($B_{\rm F450W}$, $V_{\rm F606W}$, and $I_{\rm F814W}$ for the 53W002 field;
$B_{\rm F450W}$ and $I_{\rm F814W}$ for the parallel fields)}
\end{table}

The 145 objects from Cycle 6 (Fig. 1) are plotted in Figure 2 along with the
115 objects from P96b for comparison. Unfilled symbols below the 2-$\sigma$
lower boundary are candidates that have been rejected by inspection
({\it i.e.}, they are very bright and have diffraction spikes in the \wfpc\
images, indicating they are likely galactic stars) and/or by spectroscopy. The
median
$B_{\rm F450W}$--$V_{\rm F606W}$ color (0.60 mag) of the Cycle 6 candidates
(filled squares) is fairly consistent with that of the Cycle 5 objects (0.57
mag, filled circles), seven of which have thus far been spectroscopically
confirmed (triangles). The histogram of Figure 3 shows that, for the most
part, the photometry of the field objects from all four sets of data agree
with each other. The calculated mean of the residuals of the four fields
is only (--0.07$\pm$0.01),
giving us confidence in the photometry despite the different observing times
and pointings, and the somewhat different exposure times and filter sets.
The photometry for all candidates, along with coordinates, Lyman-$\alpha$
equivalent widths, half-light radii, and redshifts (when known) are given in
Table 2.

\begin{figure}
\vspace*{-5cm}
\psfig{file=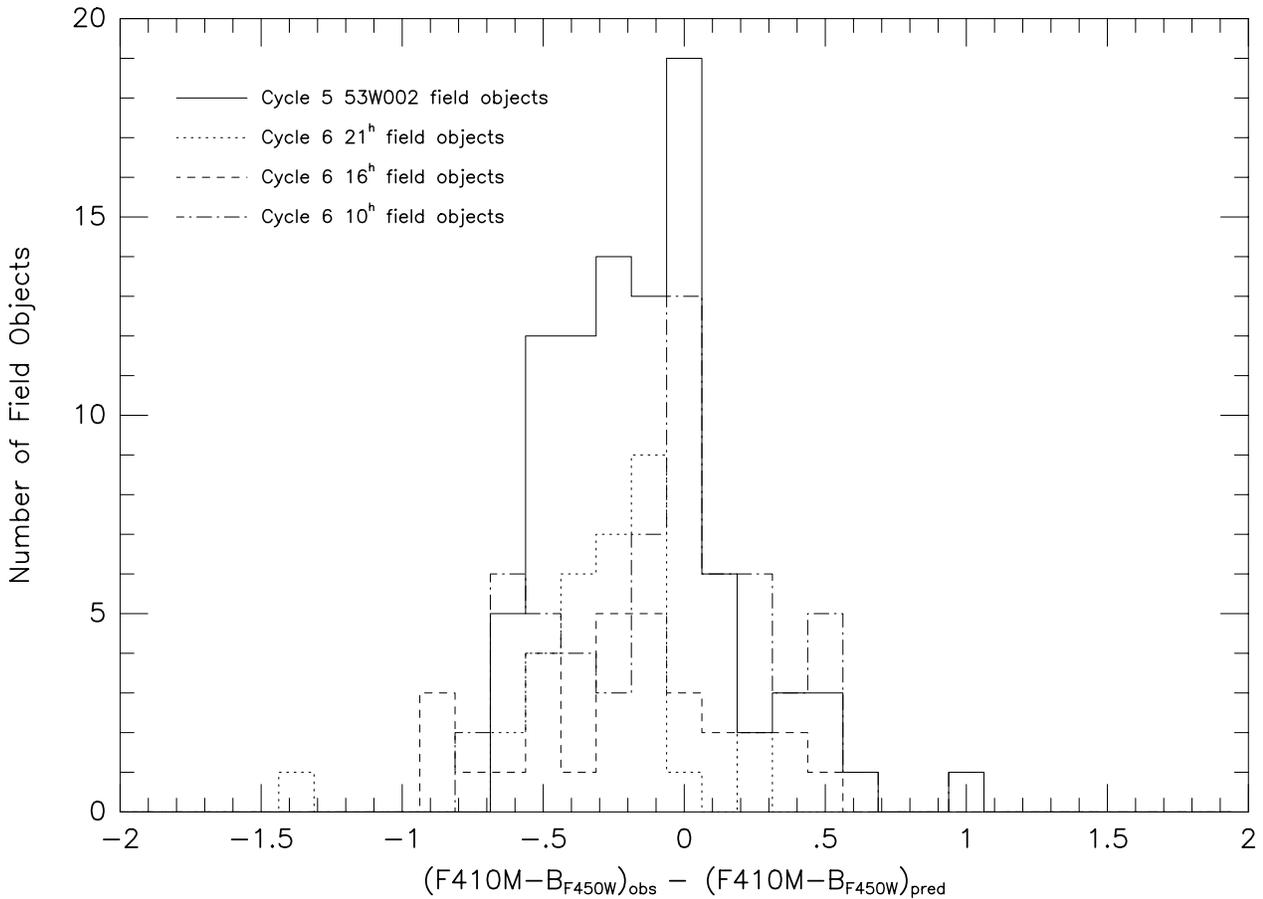,width=15cm}
\caption{Histogram of the deviations of the field objects (F410M--F450W)$_{obs}$
from the power-law line (F410M--F450W)$_{pred}$ for the
four separate sets of data. The \ztpf\ candidates
were not included in the calculations, so that an indication of the overall
consistency of the photometry could be represented. It can be seen that they
agree extremely well with each other, despite the various exposure times
and pointings.}
\end{figure}

\subsection{Sizes}

Effective radii were determined for all the candidates following the method
used in P96b, in which concentric elliptical isophotes, allowed to
vary in ellipticity and position angle, were fit to the two-dimensional object
images (see also Windhorst \etal 1994a,c; Odewahn \etal 1996; Mutz \etal 1997;
Schmidtke \etal 1997 for details). The resulting size distribution is
illustrated in Figure 4{\it a}, which also contains the Cycle 5 F410M
candidates (shaded) for comparison. The new
candidates are similarly compact, with typical $r_e$\cle 0\farcs 2, which is
smaller than the typical \wfpc\ galaxy scale lengths at these faint magnitudes
(\cle 0\farcs 3--0\farcs 4 at $B$\cle 27, Odewahn \etal 1996). Several of the
Cycle 6 objects fall into the $r_e$=0\farcs 3--0\farcs 35 bin, possibly due to
spurious detections that are not \ztpf\ objects, or (more likely) due to the
inherently lower
signal-to-noise of these images and the corresponding difficulty in determining
the proper sky level. Windhorst, Van Heerde, \& Katgert (1984) showed that
image noise can lead to an overestimate of measured sizes by
$\sim$0.2--0.3 dex. It is also possible that some fraction of the \ztpf\
candidates are as large as $r_e$=0\farcs 3.

\begin{figure}
\psfig{file=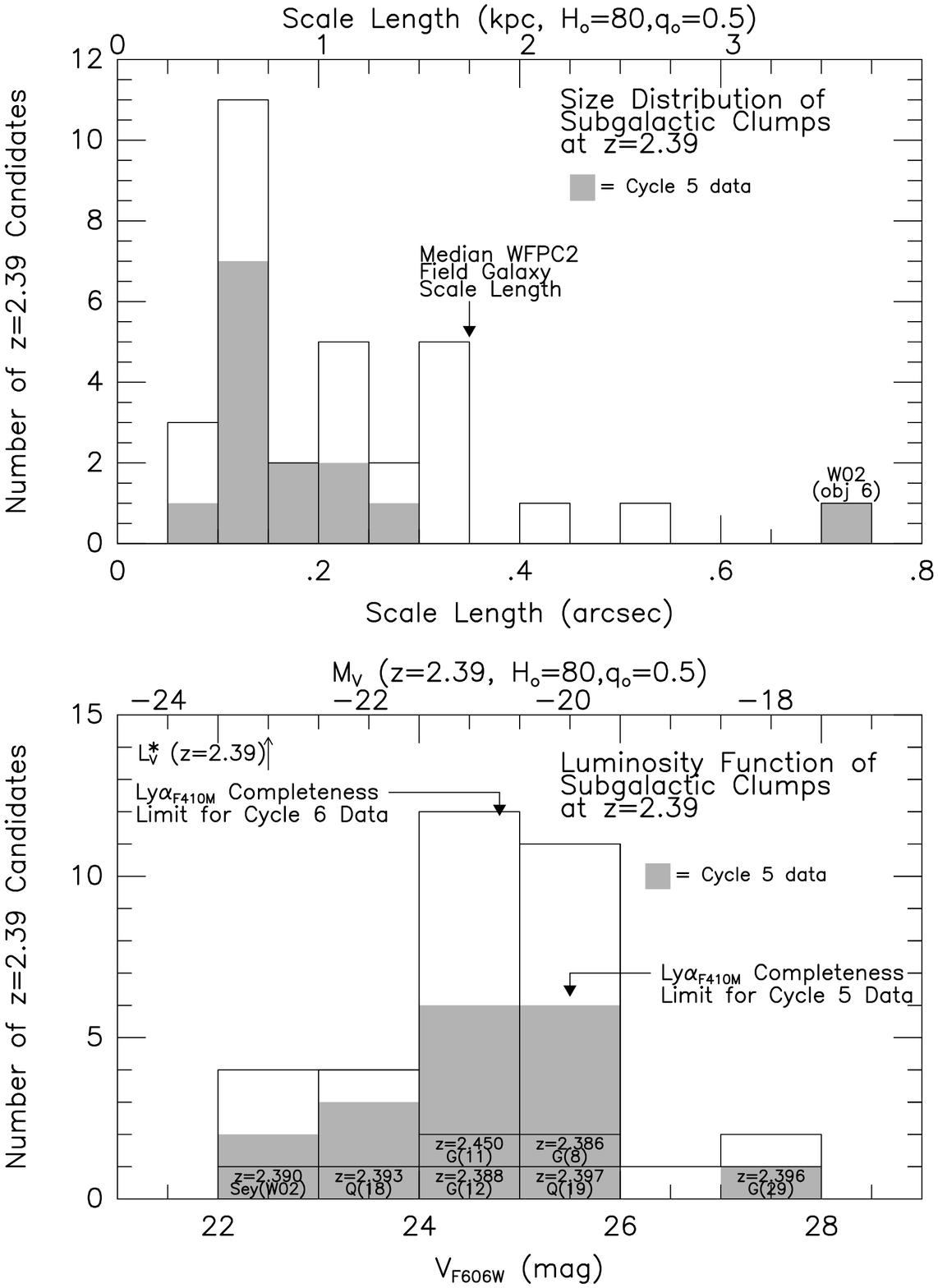,width=14.5cm}
\caption{({\it a}) Size distribution for the \ztpf\ candidates. Except for a
few extra objects in the 0\farcs 30 -- 0\farcs 35 bin, the Cycle 6 data
generally agree with that of the 53W002 field, in that most candidates are very
compact ($r_e <$ 0\farcs 2), ({\it b}) Luminosity distribution for the \ztpf\
candidates in $V_{\rm F606W}$. An absolute magnitude scale is given along the
top axis, and the (evolving) value for $L^*$ is indicated by the arrow.
In both panels, the Cycle 5 data are represented by the shaded regions, and
the Cycle 6 data by the white area. The agreement among the Cycles 5 and 6 data
sets is very good. (Note that $q_{\rm o}$=0.5 and $H_{\rm o}$=80 are used
here for direct comparison to the P96b data.)}
\end{figure}

\subsection{Luminosities}

The luminosity distribution of all the candidates from Cycles 5 (shaded) and 6
is shown in Figure 4{\it b}. Absolute magnitudes were determined from the
observed F606W magnitudes using the mean redshift ($z$=2.39) of the
confirmed objects and $K$-corrections from a galaxy SED of Bruzual \&
Charlot (1993) as in Windhorst \etal (1991). $K$-band magnitudes for several
of the candidates from recent IRTF observations of the 53W002 field (Keel \etal
1998) show a typical starburst spectrum, fairly flat in $F_{\lambda}$, so we
can do the $K$-corrections without any additional assumptions. It can be seen
in Figure 4{\it b}
that all data sets contain mostly sub-L$^*$ objects, assuming they are all at
\ztpf. In particular, a majority of the candidates have M$_{\rm V}\simeq
-20.5$ to --21.5 mag, despite the different completeness limits imposed by the
shallower F410M exposures (as indicated by the dotted arrows at the right of
Figure 2). The value for L$^*$ (M$_{\rm V}\simeq -23$ mag) was
calculated by assuming that there would have been $\sim$2 mag of stellar
evolution since \ztpf\ for a typical starburst $\sim$0.3--0.5 Gyr earlier,
as the unreddened blue colors of the candidates suggest (Bruzual \& Charlot
1993). The adopted $\sim$2 mag of luminosity evolution has an uncertainty of
$\simeq$0.50--0.75 mag and comes from the models used in Windhorst \etal
(1991) for 53W002. As described in P96b, a point source was subtracted from
three of the Cycle 5 candidates that contain an AGN based on their spectra
before being added to the histogram. No attempt was made to fit point source
contributions to any candidates without spectroscopically known AGN.

\section{Discussion}

\subsection{Galaxy Formation at High Redshift}

In models of hierarchical galaxy formation, the luminous galaxies we see today
were built up through the repeated merging of smaller protogalactic pieces.
A significant number of such sub-galactic clumps may have been found at \ztpf\
by P96b in a field surrounding the radio galaxy 53W002 at $z$=2.390. There
has been a recent series of reports of very high redshift, ``normal''
star-forming galaxies with possibly greater scale lengths and higher
luminosities ({\it e.g.}, Hu \& McMahon 1996; Madau \etal 1996; Steidel \etal
1996a and 1996b; Yee \etal 1996; Francis, Woodgate, \& Danks 1997; Lowenthal
\etal 1997; Trager \etal 1997). The Steidel \etal (1996a,b) Madau \etal
(1996), and Lowenthal \etal (1997) galaxies were found through their
$U$-band or $B$-band ``dropout,'' caused by the Lyman break in their spectra,
and not via emissions lines, which may explain why they may be more luminous
than the \ztpf\ clumps of P96b ({\it i.e}, a significant continuum detection
was necessary to find them). In addition, only about 40--50\% of these
``dropout'' galaxies are Lyman-$\alpha$ emitters, with the rest having either
no Lyman-$\alpha$ emission or significant Lyman-$\alpha$ in absorption. The
relatively stronger Lyman-$\alpha$ emitters like the sub-galactic clumps of
P96b and this paper, and those reported by Hu \& McMahon (1996) and
Francis, Woodgate, \& Danks (1997) were found likely because they
have had fewer generations of O stars, and so fewer supernovae to produce
significant dust to destroy the Lyman-$\alpha$ light.

These findings suggest that galaxies formed over a large
range of redshift rather than at one special time, and possibly
according to several different formation scenarios. If a large
percentage of galaxies did in fact form hierarchically, we would
expect to find their building blocks at high redshifts ($z$\cge 2), consistent
with the large number of star-forming objects currently being found at such
high redshifts. Many studies seem to point toward a merger rate
that was higher in the past, following $\propto$(1+$z$)$^m$ with
$m\simeq 2-4$ (Burkey \etal 1994; Carlberg, Pritchet, \& Infante 1994; Yee \&
Ellingson 1995; Patton \etal 1997) and suggests that there should have been
a higher space density of sub-galactic clumps at earlier epochs in order
to produce the observed local luminosity function. Taken together with the
fact that most luminous early-type galaxies appear to have been ``in place'' by
$z\sim 1$ (Mutz \etal 1994; Driver \etal 1995a; Lilly \etal 1995; Odewahn
\etal 1996; Heyl \etal 1997; Driver \etal 1998), these two points imply that a
large fraction of
today's luminous ellipticals and early- to mid-type spirals formed at $z$\cge
1, possibly from the merging of such sub-galactic sized objects. Indeed,
it has been shown that disks can be regenerated (or generated) after such
mergers, as gas remains bound to the system and falls back in (Hibbard \&
Mihos 1995; Tissera, Lambas, \& Abadi 1997).

As shown in P96b, \hst\ provides the resolution and sensitivity
necessary for a search for such populations of faint compact objects. The
fortunate existence of a medium-band \wfpc\ filter centered at 4100\AA\
(Lyman-$\alpha$ at \ztpf) allowed the discovery of the group of 17 candidate
\ztpf\ objects within $\sim$1 Mpc$^2$ reported by P96b. A similar finding,
using infrared filters to trace redshifted
H$\alpha$, has been reported by Malkan, Teplitz, \& McLean
(1996) at $z\simeq 2.5$. The Cycle 6 images presented in this paper
used the same medium-band \wfpc\ filter (F410M) to extend these results.
Given that the Cycle 6 observations are $\sim$50--60\% as deep as the original
53W002 observations and the luminosity function is quite steep at this
redshift, we would expect to find $<8-10$ \ztpf\ candidates in any
random part of the sky if the candidates seen around 53W002 are indeed part of
an isotropic widespread population. Here we assume that
sensitivity roughly scales as the square root of the exposure time and that
the 53W002 field is a representative high-redshift field. If, for some reason,
possibly because of the existence of the weak radio source and/or because of
the existence of several AGN surrounding 53W002 (Windhorst, Keel, \& Pascarelle
1998), this field is not
representative at \ztpf, then we should find very few (if any) significant 
\ztpf\ candidates in any of the random field searches. However, we found three
candidates in the 21$^{\rm h}$ field, 11 in the 16$^{\rm h}$ field, and four in
the 10$^{\rm h}$ field (see Fig. 1).

\subsection{Structure at \ztpf}

Currently, there are seven spectroscopically confirmed out of the 17 total
\ztpf\ candidates in the 53W002 field with three negative confirmations (P96b;
Armus \etal 1998). This implies a $\sim$70\% success rate for the medium-band
method of finding Lyman-$\alpha$ emitting candidates at high redshift. To
fairly compare the 53W002 field to the three Cycle 6 parallel fields, we must
take into account the relative depths of the observations and the probable
steepness of the \ztpf\ luminosity function. We consider all four fields only
down to the F410M completeness limit of the {\em shallowest} parallel
observation (the 21$^{\rm h}$ field at $m_{\rm F410M}$=25 mag). The 53W002
field has ten candidates brighter than this magnitude limit, and assuming a
$\sim$70\% success rate, this yields seven \ztpf\ objects. The Cycle 6 data,
for which we have
not yet attempted any spectroscopy, will yield one object (out of the two
candidates above the completeness limit) from the 21$^{\rm h}$ field, eight
objects (out of the 11 candidates above the completeness limit) from the
16$^{\rm h}$ field, and two objects (out of the three candidates above the
completeness limit) from the 10$^{\rm h}$ field. Clearly, the 16$^{\rm h}$
field is consistent with the 53W002 field, while the other two fields have a
lower density. The most striking aspect of these data is the significant
difference in the numbers of candidates from these four fields after correcting
them to the depth of the 21$^{\rm h}$ field. For any two such fields, the
significance of the different number densities can be represented by:
\begin{displaymath}
\frac{|N_1-N_2|}{\sqrt{N_1+N_2}}
\end{displaymath}
This calculation, in which $N_1$ and $N_2$ are the numbers of candidates in
each field, indicates that a 2$\sigma$ difference exists between the 21$^{\rm
h}$ (or 10$^{\rm h}$) field and 53W002 (or 16$^{\rm h}$) field. There are two
possible
reasons for this discrepancy. First, it may be an indication that the 53W002
and 16$^{\rm h}$ fields are {\it not} representative of high-redshift fields.
This would mean that they may be a $\sim$2$\sigma$ statistical fluctuation
with respect to other fields, based on the numbers estimated above. Second,
as noted in P96b, the confirmed 53W002 candidates show a remarkably small group
velocity dispersion
($\simeq 286$ km/s), despite the much larger width of the F410M filter. We
could have seen Lyman-$\alpha$ emission from objects in the entire redshift
range $z\simeq 2.28 - 2.45$, although some dropoff in transmission would occur
for $z$\cle 2.30 and $z$\cge 2.42. This
implies that the sub-galactic clumps may have existed to some extent in groups
or clusters at high redshift, or else we should have seen a larger number of
objects at other redshifts ($z \neq 2.391 \pm 0.004$) inside the F410M filter. 
Currently, we have found only one object (out of seven) which has a highly
discrepant redshift ($z$=2.451), surprisingly far out into the red wing of the
filter. Some level of clustering, therefore, can explain the statistically
marginal variation in density of the candidates from field to field. Indeed,
such a structure may have also been discovered at $z\simeq 3.0$ by Steidel
\etal (1998), lending further evidence that large-scale structure may have
existed at high redshifts.

The space density of the \ztpf\ candidates from P96b is $\sim$0.038 Mpc$^{-3}$,
assuming that Lyman-$\alpha$ lies somewhere within the F410M filter and a
$\sim$70\% success rate (= 7 objects out of 10, if we consider candidates only
down to the limiting F410M magnitude of the shallowest parallel observation).
Under the same assumptions,
we find space densities of $\sim$0.0054 Mpc$^{-3}$, $\sim$0.043 Mpc$^{-3}$, and
$\sim$0.011 Mpc$^{-3}$ for the 21$^{\rm h}$, 16$^{\rm h}$, and 10$^{\rm h}$
fields respectively. One can see that the 53W002 and 16$^{\rm h}$
fields could be a factor of 3--4 times more dense than the other two Cycle 6
fields. Unlike the 53W002 field,
the 16$^{\rm h}$ field was not targeted around an {\it a priori}
known density fluctuation such as that which might be expected in the field
surrounding the weak radio galaxy 53W002. It is interesting to note that the
10$^{\rm h}$ field,
which has a comparatively lower space density of candidates, 
was imaged in parallel to an FOC observation of the lensed object IRAS
10214+4721 at $z$=2.30, so that any associated large-scale structure would
be just within our sensitivity range.

The variation in number density among the Cycle 5 and 6 fields, if significant,
is indeed suggestive of the existence of some kind of structure ({\it e.g.},
groups, clusters, or ``sheets'' at $z\simeq 2.4$). We appear to have hit such a
structure in the 53W002 and 16$^{\rm h}$ fields, but are perhaps looking ``in
between'' any such major structures in the 21$^{\rm h}$ and 10$^{\rm h}$
fields. The small velocity dispersion of the confirmed 53W002 field objects
implies that we may be viewing such a sheet more ``face-on'' than ``edge-on,''
since the velocity dispersion would likely be much higher if a, {\it e.g.},
\cge 128 Mpc structure ({\it cf.}, Broadhurst \etal 1990; Landy \etal 1996;
Le F\`evre \etal 1994) was viewed edge-on.
A picture is beginning to develop here
which is quite consistent with that of RHS97, in which luminous galaxies at
$z$=0 are broken up into several individual clumps at higher redshifts and are
embedded in sheet- or ribbon-like structures, often lying along filaments on
these sheets. These sheets appear to be group or sub-cluster size, and may
represent the preferred environment for the formation of high-redshift objects.
Even the somewhat higher luminosity objects of Steidel \etal (1996a,b), Madau
\etal (1996), and Lowenthal \etal (1997) seem to show hints of this structure,
with several pairs and triplets having very similar redshifts (three of five
redshifts of star-forming Lyman-limit ``drop-out'' galaxies in the HDF from
Steidel \etal (1996a) are at $z\sim 2.8$). Also, there is the
significant peak in $N(z)$ at $z\simeq 3.0$ over an 11\arcmin$\times$8\arcmin
area on the sky recently reported by Steidel \etal (1998).

\subsection{The Luminosity Function of \ztpf\ Candidates}

We attempt to construct a LF for the \ztpf\ candidates from all four \wfpc\
fields based on their F450W luminosities. The \wfpc\ synthetic magnitudes were
converted to AB magnitudes and number densities were calculated using the
comoving volume defined by the redshift range of the F410M passband and the
angular size of the \wfpc\ field at \ztpf. To properly represent the fainter
luminosity bins, in which the shallower parallel fields become incomplete, we
omit candidates from fields that are not fully complete in any given
luminosity bin. The corresponding volume used in the number density
calculation is decreased accordingly, and the result is a fair indication of
the luminosity function of Lyman-$\alpha$ emitting candidates at \ztpf, if the
53W002 field is representative at this redshift.
We plot these data in Figure 5 as the filled triangles.

\begin{figure}
\vspace*{-5cm}
\psfig{file=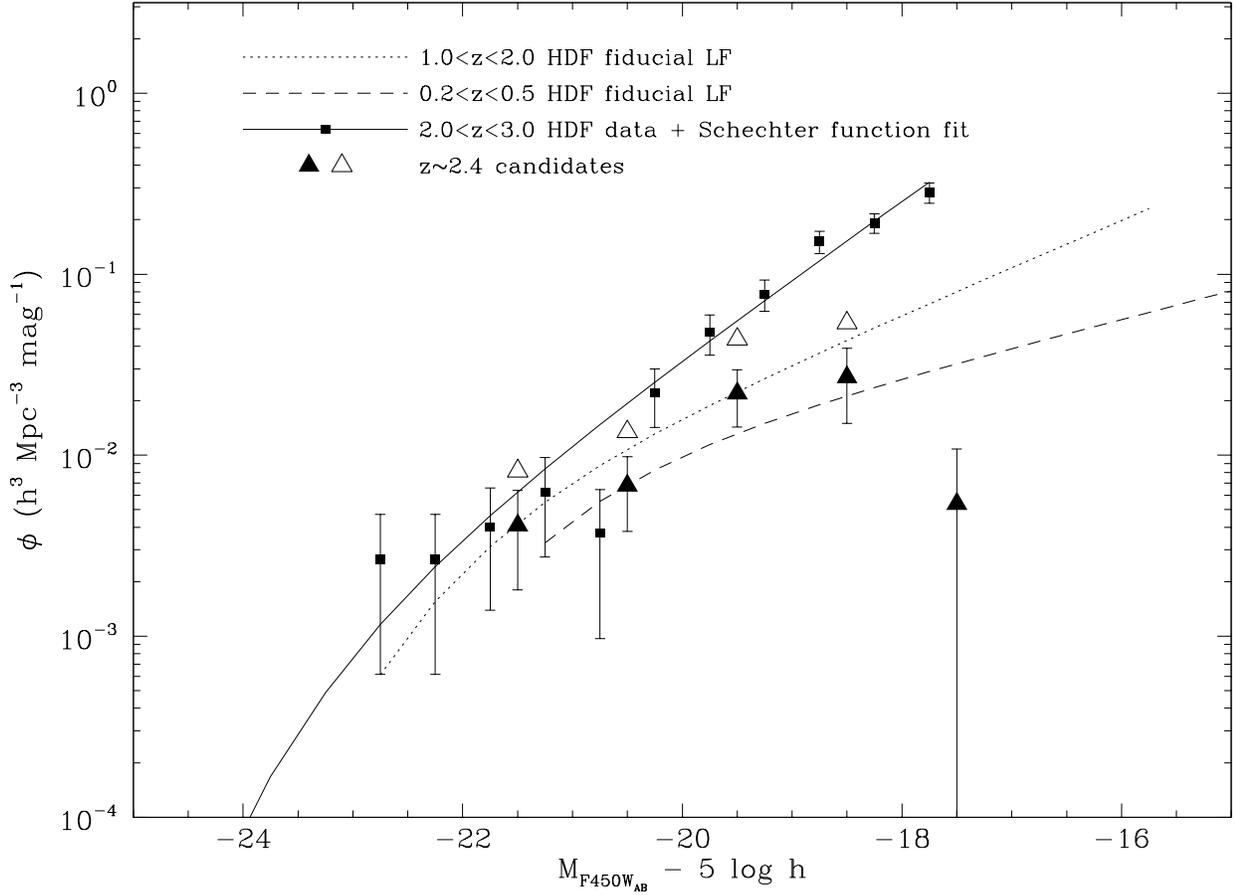,width=13cm}
\caption{Luminosity function of the Cycle 6 data (filled triangles) plotted
with the SLY97 $z$=2--3 HDF data. The data were converted to AB magnitudes
according to SLY97, and number densities were calculated using the comoving
volume defined by the redshift range of the F410M passband and the angular
size of the \wfpc\ field at \ztpf. The solid line is a Schechter luminosity
function fit to the HDF data, represented by the filled squares. Dashed and
dotted lines are fiducial LFs from other redshift bins of the SLY97 data.
Our points are slightly lower than the SLY97 data, but the results of Steidel
\etal (1996a,b), Lowenthal \etal (1997), and Trager \etal (1997) show that as
much as 50\% of their $z$=2--3 objects lack significant Lyman-$\alpha$
emission. We corrected our data for this by moving our points up by a small
amount ($\sim$0.3 dex), as indicated by the empty triangles.
}
\end{figure}

For comparison, we include in Figure 5 a Schechter LF fit (solid curve) to the
$z$=2--3 HDF field galaxy data, represented by the filled squares, from
Sawicki, Lin, \& Yee (1997, hereafter SLY97).
SLY97 find from photometric redshifts a LF for galaxies in the HDF that changes
with redshift, showing both a brightening and steepening of the HDF LF with
increasing redshift up to $z\sim 3$. Their data are in agreement with the LF of
deep galaxy redshift surveys at $z$\cle 1, {\it e.g.}, the CFRS (Lilly \etal
1995, 1996; LeFevre \etal 1996). SLY97 point out that the flattening
of the faint-end slope of the HDF luminosity function is an expected result of
hierarchical models of galaxy formation, in which merging plays an important
role in removing the sub-galactic clumps at lower redshifts.

As can be seen in Figure 5, our \ztpf\ LF is quite consistent with that of the
HDF of SLY97 for $2<z<3$ to about M$_{\rm F450W_{\rm AB}} \simeq -19.5$, below
which the completeness limit of M$_{\rm F450W_{\rm AB}} \simeq -19$ mag
imposed by our much shallower F410M observations takes over. Our
points are slightly lower than the SLY97 data, but the results of Steidel \etal
(1996a,b), Lowenthal \etal (1997), and Trager \etal (1997) show that as much
as 50\% of their $z$=2--3 objects lack significant Lyman-$\alpha$ emission.
Correcting our data for this would move our points up by a small amount
($\sim$0.3 dex). We show the corrected data points as the empty
triangles. The consistency, to within the errors, of our data with that
of SLY97 after correcting for this factor of two
supports the idea that the \ztpf\ clumps are a significant contribution to the
faint-end $z$=2--3 galaxy counts ($\sim$50\%).

It should be mentioned in this context that the scientific return from a
minimal investment of exposure time in the HDF and in the upcoming HDF South
through the F410M and F467M filters would be considerable, since all the
broadband $UBVI$ data are (or will be) in hand. This would be an excellent
direct determination of the space density of high-redshift,
Lyman-$\alpha$ emitting sub-galactic clumps.

\section{Conclusions}

We have obtained Cycle 6 \hst\ images of three random fields with the \wfpc\
F410M filter to test the idea that the sub-galactic clumps reported by P96 are
not unique to the 53W002 field (due to the presence of the radio galaxy) but are a
widespread population that exist all over the sky. We also attempt to use the
data to constrain whether or not these objects tend to be in groups
or clusters at high redshift. We have found a total of 18 candidate
Lyman-$\alpha$ emitting objects in the three parallel \hst\ fields, of which
$\sim$13 are expected to be spectroscopically confirmed if we assume the
current success rate for the 53W002 field ($\sim$70\%).
Most of these candidates appear to be as faint and compact as the Cycle 5
objects, implying that the 53W002 field is not unique, and that these faint
sub-galactic clumps were quite common at early epochs. Perhaps it is not that
surprising to find \ztpf\ objects all over the sky in light of the large
number of $z\simeq$3 objects of Steidel \etal (1998), as well as photometric
redshift samples that find peaks in the redshift distribution close to this
value (\eg, Gwyn \& Hartwick 1996 in the HDF) and point to a
peak for star formation in disk galaxies at $z\simeq 1-2$ (Pei \& Fall 1995;
Madau \etal 1996; Connolly \etal 1997). The relatively large difference in
the object counts of the three fields also hints that these objects may
preferentially exist in groups, clusters, or ``sheets,'' not unlike the
theoretical picture presented by RHS97.

Because of the existence of only two \wfpc\ filters on \hst\ suitable for
finding compact emission-line objects (the F410M and F467M filters), we were
limited in the number of redshifts we could search (\ztpf\ and $z\simeq 2.8$).
The narrow-band filters available to \wfpc\ are not wide enough to sample
more than a fraction of a single redshift ``sheet,'' if galaxies trace such
structures at high redshifts as they appear to at low redshifts (Landy \etal
1995; Cohen \etal 1996). From this perspective, the narrow \hst\ filters are
rather unsuitable for the purpose of locating larger numbers of high-redshift
Lyman-$\alpha$ emitting objects. It should also be noted that such observations
will not be possible with the Advanced Camera because it does not have the
appropriate medium-band filter set, and the field-of-view of its linear ramp
filters is too small.

The Cycles 5 and 6 data offered us a unique opportunity to examine the \ztpf\
epoch,
but similar observations with the F467M filter on \hst\ and with (hopefully)
a more extensive filter set on the Next Generation Space Telescope will surely
need to be carried out. Indeed more random observations in F410M itself are
still necessary to confirm these initial findings and improve upon our
statistics.
 
\acknowledgments

SMP acknowledges support from NASA grant NAGW-4422, NSF grant AST-9624216,
and a NASA Space Grant Fellowship. RAW and SMP
acknowledge support from \hst\ grants GO.5985.01.94A and GO.6610.01.95A,
and WCK acknowledges support from \hst\ grants GO.6610.02.95A and
GO.5985.02.94A. We would also like to thank Doug van Orsow for help in
scheduling the Cycle 6 parallel observations, Anne Cowley and David
Burstein for a careful reading of the manuscript, and Marcin Sawicki for
graciously allowing us to use the SLY97 HDF plot.

\end{document}